\documentclass[aoas,MSNbibl,nameyear,dvips]{arximspdf}
\usepackage{graphicx}
\doi{10.1214/13-AOAS640} 
\volume{7}
\issue{2}
\pubyear{2013}
\firstpage{1244}
\lastpage{1246}

\makeatletter

\newtheorem{prp}{Proposition}

\newcommand{\vbeta}{\bolds\beta}
\newcommand{\vgamma}{\bolds\gamma}
\newcommand{\vr}{{\mathbf r}}
\newcommand{\vx}{{\mathbf x}}

\makeatother

\begin{document}
\begin{frontmatter}

\title{Letter to the Editor}
\runtitle{Letter to the Editor}

\begin{aug}
\author[A]{\fnms{Yuao} \snm{Hu}},
\author[A]{\fnms{Ye} \snm{Tian}}
\and
\author[A]{\fnms{Heng} \snm{Lian}\corref{}\ead[label=e2]{henglian@ntu.edu.sg}}
\runauthor{Y. Hu, Y. Tian and H. Lian}
\affiliation{Nanyang Technological University}
\address[A]{Division of Mathematical Sciences\\
School of Physical and Mathematical Sciences\\
Nanyang Technological University\\
Singapore 637371\\
Singapore\\
\printead{e2}} 
\end{aug}

\received{\smonth{1} \syear{2013}}
\revised{\smonth{2} \syear{2013}}



\end{frontmatter}

The paper by \citet{ACG13}, \textit{Sparse least trimmed squares
regression for analyzing high-dimensional large data sets}, considered
a combination of least trimmed squares (LTS) and lasso penalty for
robust and sparse high-dimensional regression. In a recent paper
[\citet{she2011outlier}], a method for outlier detection based on a
sparsity penalty on the mean shift parameter was proposed (designated
by ``SO'' in the following). This work is mentioned in Alfons et al. as
being an ``entirely different approach.'' Certainly the problem studied
by Alfons et al. is novel and interesting. However, there is actually a
connection between the LTS approach and that of \citet{she2011outlier}.
This connection can be roughly seen from Theorem~4.1 and Proposition
4.1 of \citet{she2011outlier}, where iterative thresholding was related
to penalized regression and also to the M-estimator. In particular,
although not explicitly mentioned in \citet{she2011outlier}, from this
one can derive the close relationship between hard thresholding, $L_0$
penalty and LTS [the relationship between hard thresholding and $L_0$
penalty was mentioned on page 630 of \citet{she2011outlier}].
Given that
LTS regression is not directly posed as an M-estimator, the following
proposition can be directly shown via elementary arguments.
%
\begin{prp}
Using the notation of Alfons et al., if $(\hat{\vbeta},\hat{\vgamma
})$ is a minimizer of $\sum_{i=1}^n(y_i-\vx_i'\vbeta-\gamma
_i)^2+\lambda_1\|\vbeta\|_1+\lambda_2\|\vgamma\|_0$ and $\|\vgamma
\|_0=n-h$, then $\hat{\vbeta}$ is the minimizer of $\sum_{i=1}^h(\vr
^2(\vbeta))_{i:n}+\lambda_1\|\vbeta\|_1$.
\end{prp}

\begin{pf}
Obviously we have $\hat\gamma_i=y_i-\vx_i'\hat {\vbeta}$ if
$(y_i-\vx_i'\hat{\vbeta})^2>\lambda_2$ and $\hat\gamma_i=0$ if
$(y_i-\vx_i'\hat{\vbeta})^2<\lambda_2$. Thus, we can profile out
$\vgamma$ and get exactly the LTS problem.
\end{pf}

The result above says that a solution of SO is a solution of some LTS
problem and, thus, the set of solutions that can be obtained by SO (by
varying $\lambda_1$ and $\lambda_2$) is a subset that can be obtained
by LTS (by varying $\lambda_1$ and $h$). Obviously, if for any fixed
$\lambda_1$ and $h\ge n/2$, we can make $\|\vgamma\|_0=n-h$ by
choosing an appropriate value for $\lambda_2$, then the two will be
the same. Numerically, we do find occasionally some values of $n-h$
cannot be obtained by $\|\vgamma\|_0$. In the numerical example below
with sample size $n=59$, $h=45$ ($25\%$ trimmed) can be achieved in
both cases.

\begin{figure}

\includegraphics{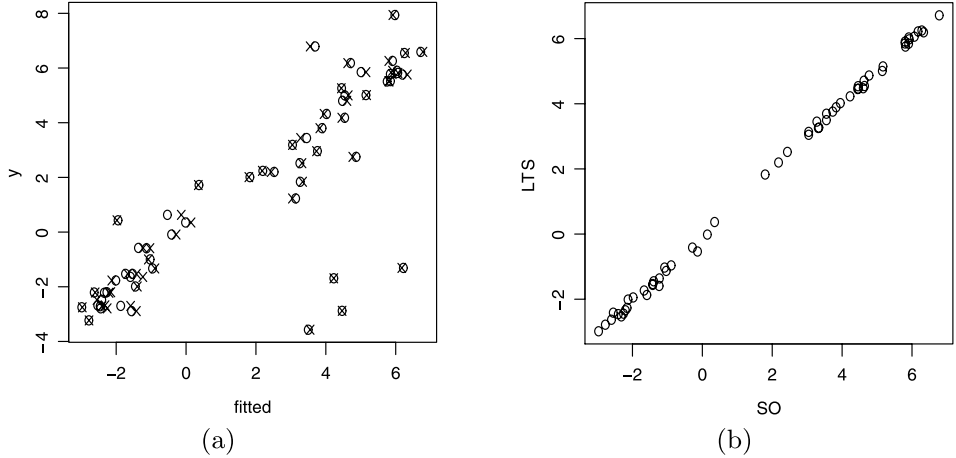}

\caption{Results for the NCI data set
with $n=59$ and $p=10$ of the 22,283 genes with the highest Spearman
rank correlation coefficients. \textup{(a)} Observed responses versus fitted
responses, where results from LTS are
denoted with the character ``$\mathrm{x}$'' and results from ``SO'' points are
denoted with the
character ``\textup{o}.'' \textup{(b)} Fitted responses from LTS versus those from SO.}
\label{fig:1}
\end{figure}

We use the same NCI-60 data to illustrate the similarities between the
two approaches. Working with the whole data ($n=59, p=22\mbox{,}283)$ using the
R package \textit{robustHD} on our desktop PC causes memory problems.
Even with $p=1000$ the program is quite slow (for both approaches). So
we use only a small number of genes just to illustrate the similarities
of the two approaches. We select $p$ genes with the largest Spearman
correlations with the response. We first use $p=10$ and $\lambda_1=0$
to avoid the complications brought about by the lasso penalty. SO is
implemented by initializing with $\vgamma=0$ and iteratively estimates
$\vbeta$ (by OLS) and $\vgamma$ (by hard thresholding). We use the
default setting with $h=45$ ($25\%$ trimmed). For SO, we set $\lambda
_2=1.34$ which results in $\|\vgamma\|_0=14$. The fitted response
values of the two approaches are shown in Figure \ref{fig:1},
demonstrating their similarity.

\begin{figure}

\includegraphics{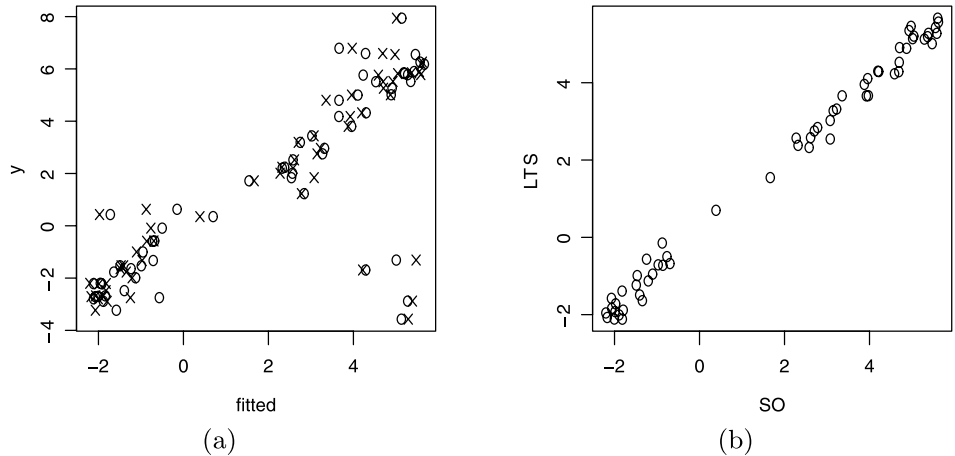}

\caption{Results for the NCI data set
with $n=59$ and $p=500$ of the 22,283 genes with the highest Spearman
rank correlation coefficients. \textup{(a)} Observed responses versus fitted
responses, where results from LTS are
denoted with the character ``$\mathrm{x}$'' and results from ``SO'' points are
denoted with the
character ``\textup{o}.'' \textup{(b)} Fitted responses from LTS versus those from SO.}
\label{fig:2}
\end{figure}

As a second illustration, we use $p=500$ genes. We find that BIC values
for the LTS approach decrease as $\lambda_1\rightarrow0$, possibly
because we picked genes with the largest correlations with the
response. So we just manually set the parameter for the lasso penalty
to be $0.1$ in the $\mathit{sparseLTS}()$ function of the \textit{robustHD}
package. Based on equation (1.4) in \citet{ACG13}, this
actually should correspond to $\lambda_1=h\times0.1=4.5$. However,
this value of $\lambda_1$ was too large for the SO implementation and
resulted in $\vbeta=0$. Thus, we perform a two-dimensional search to
find the values of $(\lambda_1,\lambda_2)$ that produce a similar
solution (in particular, with the same number of outliers), and finally
find $\lambda_1=0.26, \lambda_2=1.44$. The fitted response values for
the two approaches are shown in Figure \ref{fig:2}. There is a larger
difference between the two approaches compared to Figure \ref{fig:1}.
The difference might be due to different initialization methods,
numerical errors or convergence issues. We also note that the
initialization method used for penalized LTS is random and multiple
executions of the same function in \textit{robustHD} will produce
slightly different results.
%

\printaddresses


\begin{thebibliography}{2}

\bibitem[\protect\citeauthoryear{Alfons, Croux and Gelper}{2013}]{ACG13}
\begin{barticle}[mr]
\bauthor{\bsnm{Alfons},~\bfnm{Andreas}\binits{A.}},
  \bauthor{\bsnm{Croux},~\bfnm{Christophe}\binits{C.}} \AND
  \bauthor{\bsnm{Gelper},~\bfnm{Sarah}\binits{S.}}
(\byear{2013}).
\btitle{Sparse least trimmed squares regression for analyzing
high-dimensional large data sets}.
\bjournal{Ann. Appl. Stat.}
\bvolume{7}
\bpages{226--248}.
\bptok{imsref}%
\end{barticle}
\endbibitem

\bibitem[\protect\citeauthoryear{She and Owen}{2011}]{she2011outlier}
\begin{barticle}[mr]
\bauthor{\bsnm{She},~\bfnm{Yiyuan}\binits{Y.}} \AND
  \bauthor{\bsnm{Owen},~\bfnm{Art~B.}\binits{A.~B.}}
(\byear{2011}).
\btitle{Outlier detection using nonconvex penalized regression}.
\bjournal{J. Amer. Statist. Assoc.}
\bvolume{106}
\bpages{626--639}.
\bid{doi={10.1198/jasa.2011.tm10390}, issn={0162-1459}, mr={2847975}}
\bptok{imsref}%
\end{barticle}
\endbibitem

\end{thebibliography}
\end{document}